\documentstyle[12pt]{article}

\catcode`\@=11
\def\marginnote#1{}
\newcount\hour
\newcount\minute
\newtoks\amorpm
\hour=\time\divide\hour by60
\minute=\time{\multiply\hour by60 \global\advance\minute
by-\hour}\edef\standardtime{{\ifnum\hour<12
\global\amorpm={am}%
        \else\global\amorpm={pm}\advance\hour by-12 \fi
        \ifnum\hour=0 \hour=12 \fi
        \number\hour:\ifnum\minute<10
0\fi\number\minute\the\amorpm}}
\edef\militarytime{\number\hour:\ifnum\minute<10
0\fi\number\minute}

\def\draftlabel#1{{\@bsphack\if@filesw {\let\thepage\relax
   \xdef\@gtempa{\write\@auxout{\string
      \newlabel{#1}{{\@currentlabel}{\thepage}}}}}\@gtempa
   \if@nobreak \ifvmode\nobreak\fi\fi\fi\@esphack}
        \gdef\@eqnlabel{#1}}
\def\@eqnlabel{}
\def\@vacuum{}
\def\draftmarginnote#1{\marginpar{\raggedright\scriptsize\tt#1}}
\def\draft{\oddsidemargin -.5truein
        \def\@oddfoot{\sl preliminary draft \hfil
        \rm\thepage\hfil\sl\today\quad\militarytime}
        \let\@evenfoot\@oddfoot \overfullrule 3pt
        \let\label=\draftlabel
        \let\marginnote=\draftmarginnote

\def\@eqnnum{(\theequation)\rlap{\kern\marginparsep\tt\@eqnlabel}%
\global\let\@eqnlabel\@vacuum}  }

\def\numberbysection{\@addtoreset{equation}{section}
        \def\theequation{\thesection.\arabic{equation}}}

\def\underline#1{\relax\ifmmode\@@underline#1\else
 $\@@underline{\hbox{#1}}$\relax\fi}

\catcode`@=12
\relax

\def\fin{\end{document}}
\def\beq{\begin{equation}}
\def\eeq{\end{equation}}
\def\bea{\begin{eqnarray}}
\def\eea{\end{eqnarray}}

\def\cg{{\cal G}}
\def\cgh{{\hat {\cal G}}}
\def\ch{{\cal H}}
\def\chh{{\hat {\cal H}}}

\def\Gcc#1 {{\cal G}^{[ #1 ]}_\|}
\def\Gcp#1 {{\cal G}^{[ #1 ]}_\bot}
\def\Cco#1 {{\cal C}^{[ #1 ]} }

\def\rf#1{(\ref{eq:#1})}
\def\lab#1{\label{eq:#1}}
\def\nonu{\nonumber}
\def\br{\begin{eqnarray}}
\def\er{\end{eqnarray}}
\def\be{\begin{equation}}
\def\ee{\end{equation}}

\def\lb{\lbrack}
\def\rb{\rbrack}

\def\({\left(}
\def\){\right)}

\newcommand{\ct}[1]{\cite{#1}}

\relax

\def\Tr{\mathop{\rm Tr}}

\def\a{\alpha}
\def\b{\beta}

\def\d{\delta}

\def\l{\lambda}

\def\o{\over}
\def\om{\omega}

\def\pr{\prime}
\def\ra{\rightarrow}

\def\cf{{\cal F}}

\def\lie{{\cal G}}

\def\f#1#2#3 {f^{#1#2}_{#3}}

\def\win1{{\sf w_{1+\infty}}}

\def\Win1{{\sf W_{1+\infty}}}

\def\rlx{\relax\leavevmode}
\def\inbar{\vrule height1.5ex width.4pt depth0pt}
\def\IZ{\rlx\hbox{\sf Z\kern-.4em Z}}
\def\IR{\rlx\hbox{\rm I\kern-.18em R}}
\def\IC{\rlx\hbox{\,$\inbar\kern-.3em{\rm C}$}}
\def\IN{\rlx\hbox{\rm I\kern-.18em N}}
\def\IO{\rlx\hbox{\,$\inbar\kern-.3em{\rm O}$}}
\def\IP{\rlx\hbox{\rm I\kern-.18em P}}
\def\IQ{\rlx\hbox{\,$\inbar\kern-.3em{\rm Q}$}}
\def\IF{\rlx\hbox{\rm I\kern-.18em F}}
\def\IG{\rlx\hbox{\,$\inbar\kern-.3em{\rm G}$}}
\def\IH{\rlx\hbox{\rm I\kern-.18em H}}
\def\II{\rlx\hbox{\rm I\kern-.18em I}}
\def\IK{\rlx\hbox{\rm I\kern-.18em K}}
\def\IL{\rlx\hbox{\rm I\kern-.18em L}}
\def\one{\hbox{{1}\kern-.25em\hbox{l}}}
\def\0#1{\relax\ifmmode\mathaccent"7017{#1}%
        \else\accent23#1\relax\fi}

\def\NPB#1#2#3{{\sl Nucl. Phys.} {\bf B#1} (#2) #3}

\def\CMP#1#2#3{{\sl Commun. Math. Phys.} {\bf #1} (#2) #3}

\def\PLB#1#2#3{{\sl Phys. Lett.} {\bf #1B} (#2) #3}

\def\JPA#1#2#3{{\sl J. Physics} {\bf A#1} (#2) #3}

\begin{document}

\begin{titlepage}
\nopagebreak
\begin{flushright}
SWAT/93-94/44\\
US-FT/14-94\\
IFT-P.039/94\\
hep-th/9411036
\\
October 1994
\end{flushright}

\vglue 2.5  true cm
\begin{center}
{\large\bf
ORTHOGONAL DECOMPOSITION OF SOME \\AFFINE LIE ALGEBRAS \\
IN TERMS OF THEIR HEISENBERG SUBALGEBRAS}
\end{center}
\normalsize
\vskip .4in

\begin{center}
L.A. Ferreira$^{\dagger}$\footnotemark
{\footnotetext{On leave from Instituto de F\'\i sica Te\'orica,
IFT/UNESP - S\~ao Paulo-SP - Brazil.}}, D.I. Olive$^{\dagger\dagger}$ and
M.V. Saveliev$^{\dagger\dagger\dagger}$
\par \vskip .5in \noindent
$^{\dagger}$Facultad de F\'\i sica \\
Universidade de Santiago de Compostela\\
15706 Santiago de Compostela\\
Spain
\par \vskip .3in

\end{center}

\begin{center}

$^{\dagger\dagger}$Department of Physics\\
University of Wales, Swansea\\
Swansea, SA2 8PP, Wales\\
United Kingdom\\

\par \vskip .3in

\end{center}

\begin{center}

$^{\dagger\dagger\dagger}$Institute for High Energy Physics\\
142284, Protvino, Moscow region\\
Russia
\par \vskip .3in

\end{center}

\vfill
\begin{abstract}
\baselineskip .4 true cm
\noindent
In the present note we suggest an affinization of a theorem by Kostrikin et.al.
about the decomposition of some complex simple Lie algebras ${\cal G}$ into
the algebraic sum of  pairwise orthogonal Cartan subalgebras.
We point out that the untwisted affine Kac-Moody algebras of types $A_{p^m-1}$
($p$ prime, $m\geq 1$), $B_r, \, C_{2^m}, D_r,\, G_2,\, E_7,\, E_8$ can be
decomposed into  the algebraic sum of pairwise or\-tho\-go\-nal Heisenberg
subalgebras.  The  $A_{p^m-1}$ and $G_2$ cases are discussed in great detail.
Some   possible applications of such decompositions are also discussed.

\end{abstract}
\vfill
\end{titlepage}

\baselineskip .5 true cm

\section{Introduction}

In the early eighties A. I. Kostrikin, I. A. Kostrikin and V. A. Ufnarovskii
 proved a theorem \cite{KKU81}, see
also \cite{KKU811} and \cite{KKU82}, that the
complex Lie algebras $\cg$ of types
$A_{p^m-1}$ ($p$ prime,
$m\geq 1$), $B_r, \, C_{2^m}, D_r,\, G_2,\, E_7,\, E_8$ can be decomposed into
the algebraic sum of Cartan subalgebras, pairwise orthogonal with respect
to the Killing form; i.e.
\be
\cg = \ch_0 \, \oplus \, \ch_1 \, \oplus \, \cdots \, \oplus \, \ch_h
\lab{decom}
\ee
with
\be
\Tr \( \ch_i \, \ch_j \) = 0 \, \, \, \, \mbox{\rm for $i \neq j$},
\lab{otho}
\ee
where $h$ is the Coxeter number of $\cg$. The fact that the
number of Cartan subalgebras in the above decomposition \rf{decom} is $h+1$
accords with the well-known observation that for all semisimple algebras
$h+1 = \mbox{\rm dim $\cg$}/r,\; r=\mbox{\rm rank $\cg$}$.

Here we show that this theorem can be extended to many of the
corresponding affine Lie algebras $\cgh$, in the sense that the latter may
be decomposed  into the algebraic  sum of pairwise orthogonal Heisenberg
subalgebras. Those  subalgebras are in fact the affinization of the Cartan
subalgebras in the  corresponding decomposition of the finite--dimensional
algebras $\cg$. The problem of extending the results to the affine case
consists in identifying the gradations of the Kac-Moody algebra $\cgh$
which respect the decomposition \rf{decom}.

The integral gradations of an affine Kac-Moody algebra $\cgh$ are
provided by the grading
operators \cite{KP85}
\be
Q = \sum_{a=1}^{r} s_a \l^v_a \cdot H + N d,
\lab{gradop}
\ee
where $(s_0, s_1, \cdots ,s_r)$ is a vector of
non negative, relatively prime  integers, $\l^v_a \equiv 2
\lambda_a/\alpha_a^2$
with $\lambda_a$ and $\alpha_a$ being  the fundamental weights and simple roots
of $\cg$ respectively. Furthermore,
\be
N = \sum_{i=0}^{r} s_i \, m_i^{\psi}, \qquad
\psi = \sum_{a=1}^{r}m_a^{\psi}\,  \alpha_a , \qquad
m_0^{\psi} \equiv 1
\lab{grad}
\ee
and $\psi$ is the maximal root of $\cg$. Two gradations are counted as
equivalent if the corresponding vectors $(s_0,s_1,\cdots ,s_r)$ and $(s'_0,
s'_1, \cdots s'_r)$ are related by a symmetry of the extended Dynkin diagram
 of $\cg$.

Given a finite--dimensional complex simple Lie algebra $\cg$ with
decomposition \rf{decom}, we look
for a gradation such that the corresponding affine Kac--Moody algebra $\cgh$
can be written as
\be
\cgh = \chh_0 \, \oplus \, \chh_1 \, \oplus \, \cdots \, \oplus \, \chh_h,
\lab{decomkm}
\ee
where each subspace $\chh_i$ is a Heisenberg subalgebra of $\cgh$  with
generators being eigenvectors of the corresponding grading operator $Q$. In
addition, under the restriction of $\cgh$ to $\cg$, the elements of $\chh_i$
become those of $\ch_i$.

For the case of the homogeneous gradation it is quite trivial to make
the extension of the decomposition \rf{decom} with the orthogonality
property \rf{otho} to the affine case. In this case
$s_0=1$ and $s_a =0$ for $a=1,2\cdots ,r$, and so $Q=d$. This gradation is
insensitive to the structure of the root system of $\cg$  and therefore it does
not matter what linear combinations of the Weyl--Cartan basis elements the
generators of $\ch_i$ are. We construct the generators of $\chh_i$ by
replacing in $\ch_i$ the elements $H_a$ by $H_a^n$ and $E_{\a}$ by
$E_{\a}^n$, with $n$
being the integral eigenvalues of $d$; in other words $\lb d , H_a^n \rb =
n H_a^n$, and
$\lb d , E_{\a}^n \rb = n E_{\a}^n$. The orthogonality condition is preserved
since if, say $S$ and $T$ are two orthogonal elements of $\cg$, so are the
elements  $S^m$ and $T^n$ of $\cgh$; i.e., $\Tr ( S^m T^n ) = \Tr ( S T )
\, \d_{m+n,0}$.

For other gradations the extension is more delicate, and we discuss here
the case of the Lie algebras $A_{p^m-1}$ and $G_2$ only. The cases of
gradations which are not of type \rf{gradop} deserve further study. At the
end of section \ref{sec2} we comment on a special gradation of $A_2$.

\section{The case $A_{p-1}$ with $p$ prime}
\label{sec2}

In refs. \cite{KKU81}, \cite{KKU811} (see also \cite{PZ81}) the algebra
$A_{p-1}$ was decomposed as in
\rf{decom} using some special properties of the defining  representation.
Consider the $p\times p$ matrices
\be
g=\(
\begin{array}{cccccc}
0&x_1&0&0&\cdots &0\\
0&0&x_2&0&\cdots &0\\
0&0&0&x_3&\cdots &0\\
\vdots&\vdots&\vdots&\vdots&\vdots&\vdots\\
0&0&0&0&\cdots &x_{p-1}\\
x_0&0&0&0&\cdots &0
\end{array}\),
\qquad
D=\(
\begin{array}{cccccc}
1&0&0&0&\cdots &0\\
0&\om&0&0&\cdots &0\\
0&0&\om^2&0&\cdots &0\\
\vdots&\vdots&\vdots&\vdots&\vdots&\vdots\\
0&0&0&\cdots &\om^{p-2}&0\\
0&0&0&0&\cdots &\om^{p-1}
\end{array}\),
\lab{gd}
\ee
where $x_i$, $i=0,1,2\ldots p-1$ are arbitrary non vanishing constants,
$\om$ is the $p$-th root of unit ($\om =\exp (2\pi
i/p), \, \om^p=1$), and so
\be
g \, D = \om \, D \, g,  \qquad  g^p =x_0 x_1 \ldots x_{p-1}\one , \qquad
D^p=\one .
\lab{prop}
\ee
The subspaces in \rf{decom} are defined as
\br
\ch_0 &=& \{ D^k \, , \, k=1,2,3, \ldots ,p-1\};\nonu\\
\ch_{l+1} &=& \{ \( D^{l} g\)^k \, , \, k=1,2,3, \ldots ,p-1\},
\qquad l=0,1,2 ,\ldots ,p-1;
\lab{subsp}
\er
and the Coxeter number $h$ of $A_{p-1}$ equals $p$.

The  subspaces \rf{subsp} are obviously abelian. However, the linear
independency and  orthogonality of them requires $p$ to be a prime number.
Using \rf{prop} one observes that
\be
\( D^{l} g\)^k \sim  \, D^{lk} g^k,
\ee
where here, and below, the symbol $\sim $ means the two quantities are
proportional (the proportionality constant being $\om$ dependent).
Therefore, if for given $l$ and $l^{\pr}$ one has $lk = l^{\pr} k + mp$,
with $m$
integer, then $\( D^{l} g\)^k \sim \( D^{l^{\pr}} g\)^k$. Now, if $p$ is prime,
either $k$ or $(l- l^{\pr})$ will have to have $p$ in its expansion as a
product of primes. But this can not happen since they can be at most $(p-1)$.
Therefore  the subspaces \rf{subsp} are linearly independent for $p$ prime. As
for the orthogonality, consider $\Tr \( \( D^{l} g\)^k \( D^{l^{\pr}}
g\)^{k^{\pr}} \) \sim \Tr \( D^{lk+l^{\pr}k^{\pr}} g^{k+k^{\pr}}\)$. That does
not vanish for $l\neq  l^{\pr}$ only if $k+k^{\pr}= mp$ and
$lk+l^{\pr}k^{\pr}=np$, with $m$ and $n$ integers, and so for
$(l-l^{\pr})k=(n-ml^{\pr})p$, which is impossible for $p$ prime.

Now to extend the consideration to the affine case, we have to select a
gradation which respects
the decomposition written above. As we have already said, the homogenous
gradation fits our
needs. However, in this particular case the principal gradation also does, and
we find it more interesting and important for applications (see comments at
the end of this section on an additional gradation for $A_2$).

In fact the homogeneous and principal gradations are the only integer
gradations which respect the decomposition above. In order to see that, one
observes that the elements $D^lg$ are linear combinations of the step operators
for simple roots $E_{\a_a}$, $a=1,2,\ldots ,p-1$, and $E_{-\psi}$ where $\psi$
is the highest root. Now when we affinize the algebra we will need to find
integers $n_a$ and $n_b$ such that $E_{\a_a}^{n_a}$ and $E_{\a_b}^{n_b}$
have  the same grade under the grading operator
\rf{gradop}. Therefore we need  $n_b  = n_a +{s_a - s_b\o N}$. But since $s_a$
and $s_b$ are non negative it follows that $\mid s_a - s_b\mid$ is smaller than
$s_a$ and/or $s_b$. On the other hand, from \rf{grad} we see that $N$ is
greater
than  $s_a$ and $s_b$. So it is impossible to satisfy the above equality with
$n_a$ and $n_b$ integers unless
$s_a = s_b$ for all $a$ and $b$. Now we also need $E_{\a_a}^{n_a}$ and
$E_{-\psi}^{n_0}$ to have the same grade, and so $s_a + N n_a =
-\sum_{b=1}^{p-1} s_b + N n_0$ (since $\psi = \sum_{b=1}^{p-1} \a_b$).
Therefore we need $n_a = n_0 - {p s \o {(p-1)s + s_0}}$, where we used the fact
that all $s_a$'s are equal and have set $s_a \equiv s$ for any $a$. So we have
to have $s=0$ (homogeneous) or $s=s_0$ (principal).

We then consider the principal gradation and  define the following generators
of
the affine Kac--Moody algebra
$\hat{sl}(p, \IC )$:
\br
\chh^{0}_{k,pn} \equiv \om_k H_1^n + \om_k^2 H_2^n + \ldots
+ \om_k^{p-1} H_{p-1}^n +  H_0^n,
\lab{h0}
\er
with $k=1,2, \ldots ,p-1$; and also
\br
\chh^{l+1}_{1,pn+1} &\equiv& \om_l E_{\a_1}^n + \om_l^2 E_{\a_2}^n +
\om_l^3  E_{\a_3}^n + \ldots + \om_l^{p-1} E_{\a_{p-1}}^n +
 E_{\a_{0}}^{n+1};\\
\chh^{l+1}_{2,pn+2} &\equiv& \om_{2l} E_{\a_1+\a_2}^n + \om_{2l}^2
E_{\a_2+\a_3}^n +
\om_{2l}^3  E_{\a_3+\a_4}^n + \ldots \nonu\\
&+& \om_{2l}^{p-2} E_{\a_{p-2}+\a_{p-1}}^{n}
+ \om_{2l}^{p-1} E_{\a_{p-1}+\a_0}^{n+1} + E_{\a_{0}+\a_1}^{n+1};\\
\chh^{l+1}_{3,pn+3} &\equiv& \om_{3l} E_{\a_1+\a_2+\a_3}^n + \om_{3l}^2
E_{\a_2+\a_3+\a_4}^n + \om_{3l}^3  E_{\a_3+\a_4+\a_5}^n + \ldots \nonu\\
&+& \om_{3l}^{p-3} E_{\a_{p-3}+\a_{p-2}+\a_{p-1}}^{n}
+ \om_{3l}^{p-2} E_{\a_{p-2}+\a_{p-1}+\a_0}^{n+1} \nonu\\
&+& \om_{3l}^{p-1} E_{\a_{p-1}+\a_0+\a_1}^{n+1}
+ E_{\a_{0}+\a_1+\a_2}^{n+1};\\
\vdots &\vdots& \vdots \nonu\\
\chh^{l+1}_{k,pn+k} &\equiv& \om_{kl} E_{\a_1+\ldots +\a_k}^n + \om_{kl}^2
E_{\a_2+\ldots +\a_{k+1}}^n + \om_{kl}^3  E_{\a_3+\ldots +\a_{k+2}}^n + \ldots
\nonu\\
&+& \om_{kl}^{p-k} E_{\a_{p-k}+\ldots +\a_{p-1}}^{n}
+ \om_{kl}^{p-k+1} E_{\a_{p-k+1}+\ldots +\a_{p-1}+\a_0}^{n+1} + \ldots \nonu\\
&+& \om_{kl}^{p-1} E_{\a_{p-1}+\a_0+\a_1+\ldots + \a_{k-2}}^{n+1}
+ E_{\a_{0}+\a_1+\ldots +\a_{k-1}}^{n+1};\\
\vdots &\vdots& \vdots \nonu\\
\chh^{l+1}_{p-1,pn+p-1} &\equiv& \om_{(p-1)l} E_{\a_1+\a_2+\ldots +\a_{p-1}}^n
+
\om_{(p-1)l}^2 E_{\a_2+\a_3+\ldots +\a_{p-1} + \a_0}^{n+1}\nonu\\
& +&
\om_{(p-1)l}^3  E_{\a_3+\ldots +\a_{p-1}+\a_0 + \a_1}^{n+1} +
\ldots
\nonu\\
&+& \om_{(p-1)l}^{p-1} E_{\a_{p-1}+\a_{0}+\a_{1}+\ldots +\a_{p-3}}^{n+1}
+ E_{\a_{0}+\a_1+\ldots +\a_{p-2}}^{n+1},
\lab{step}
\er
with $l=0,1,2, \ldots ,p-1$. Here we have denoted $\om_k \equiv \exp (2 \pi
i k/p)$, and $\a_i$, $i=0,1,2, \ldots ,p-1$ are the simple roots of
$\hat{sl}(p, \IC )$ with $\a_0 = - \psi$; $\ \psi \equiv$ is the maximal
root of
$\hat{sl}(p, \IC )$.  The elements $H_i^n \equiv 2 \a_i \cdot H^n/\a_i^2 +
\d_{i,0} C$ with
$C$ being the centre of $\hat{sl}(p, \IC )$,
$i=0,1,2, \ldots ,p-1$ and $E_{\a}^n$ are the generators of  $\hat{sl}(p,
\IC )$ in
the Chevalley basis. They satisfy the commutation relations
\br
\lb H_a^m \, , \, H_b^n \rb &=& C {2\o \a_a^2} K_{ab} m \d_{m+n,0} \nonu\\
\lb H_a^m \, , \, E_{\pm \a}^n \rb &=& \pm K_{\a a} E_{\pm \a}^{m+n} \nonu \\
\lb E_{\a}^m \, , \, E_{-\a}^n \rb &=&  l_a^{\a} H_a^{m+n} +
C {2\o \a^2} m \d_{m+n,0} \nonu\\
\lb E_{\a}^m \, , \, E_{\b}^n \rb &=& \epsilon (\a , \b )
E_{\a + \b}^{m+n}; \, \, \, \qquad \mbox{\rm if $\a + \b$ is a root of $\lie$}
\nonu\\
\lb d \, , \, H_b^m \rb &=& m H_b^m \nonu\\
\lb d \, , \, E_{\a}^m \rb &=& m  E_{\a}^m
\lab{km}
\er
with $a,b=1,2, \ldots ,p-1$;  $K_{\a a}= 2 \a \cdot \a_a / \a_a^2 = n_b^{\a}
K_{ba}$ with $K_{ab}$ being the Cartan matrix of $\lie$;
$n_a^{\a}$ and $l^{\a}_a$ are the integers in the expansions
$\a = \sum_{a=1}^r n^{\a}_a \a_a$ and $\a /\a^2 = \sum_{a=1}^r l^{\a}_a \a_a /
\a_a^2$, and
$\epsilon (\a , \b )$ are structure constants.

Notice all the generators defined above are eigenvectors of the principal
gradation operator, see \rf{gradop},
\be
 Q_{\mbox{\rm pr.}} = \sum_{a=1}^{p-1} \l^v_a \cdot H + p d,
\ee
i.e.
\br
\lb Q_{\mbox{\rm pr.}} \, , \, \chh_{k,pn}^0 \rb &=& pn \, \chh_{k,pn}^0, \\
\lb Q_{\mbox{\rm pr.}} \, , \, \chh_{k,pn+k}^{l+1} \rb &=&
(pn+k) \, \chh_{k,pn+k}^{l+1}.
\er

We then define the subspaces in \rf{decomkm} as
\br
\chh_0 &\equiv & \{ \chh_{k,pn}^0 \, , \, k=1,2,\ldots ,p-1 \, , \,
 n \in \IZ \};\nonu\\
\chh_{l+1} &\equiv & \{ \chh_{k,pn+k}^{l+1} \, , \, k=1,2,\ldots ,p-1 \, , \,
 n \in \IZ \} \qquad l=0,1, \ldots ,p-1.
\lab{kmsubspaces}
\er

Each one of these subspaces constitutes a Heisenberg
subalgebra of $\hat{sl}(p, \IC )$. The first one, $\chh_0$, is a homogeneous
Heisenberg subalgebra and the others,  $\chh_{l+1}$, are principal ones.
The proof that this is indeed so, can be obtained by realizing the
Kac-Moody algebra \rf{km} in the following way.

Introduce the bracket
\be
\lb A(z)\, , \, B(z) \rb \equiv \lb A(z)\, , \, B(z) \rb_{\mbox{\rm o.c.}} +
C \one \oint {dz \o 2 \pi i } \Tr ( B(z) {d\, \o dz}A(z))
\lab{bracket}
\ee
where $\lb \, , \,  \rb_{\mbox{\rm o.c.}}$ is the ordinary commutator between
matrices, i.e. $\lb X\, , \, Y \rb_{\mbox{\rm o.c.}} \equiv  X Y - Y X$. The
commutation relations \rf{km} can be obtained from this bracket by writing the
Kac-Moody generators as the matrices:
\br
\( H_a^{n}\)_{ij}&\equiv& z^n (\d_{i,a}\d_{j,a} - \d_{i,a+1}\d_{j,a+1}),\nonu\\
\( H_0^{n}\)_{ij}&\equiv& z^n (-\d_{i,1}\d_{j,1} + \d_{i,p}\d_{j,p}) + C
\d_{n,0}\,\d_{ij},
\nonu\\
\( E_{\a_a}^n\)_{ij}&\equiv& z^n \d_{i,a} \d_{j,a+1}, \nonu\\
\( E_{\a_0}^n\)_{ij}&\equiv& z^n \d_{i,p} \d_{j,1}, \nonu\\
\( E_{\a_a+\a_{a+1}+\ldots +\a_{a+k}}^n\)_{ij} &\equiv& z^n \d_{i,a}
\d_{j,a+k+1}
\lab{defrep}
\er
with $i,j=1,2,\ldots p$; $a=1,2,\ldots p-1$ and $k=1,2,\ldots p-a-1$. The
negative root step operators are obtained from the positives ones by
transposing the matrices and inverting the power of $z$, i.e.
$E_{-\a}^n =\( E_{\a}^{-n}\)^{\dagger}$
(notice that for $k>l$, $\a_k + \a_{k+1}+\ldots +\a_{p-1}+\a_0+\a_1+\a_2+\ldots
+\a_l=-\a_{l+1}-\a_{l+2}-\ldots -\a_{k-1}$). In addition we have $d\equiv \one
\,  z {d\, \o dz}$.

By introducing the matrix
\be
g(z) \equiv  g\mid_{x_0=z,x_a=1} \equiv E_{\a_1}^0  + E_{\a_2}^0 + \ldots
+ E_{\a_{p-1}}^0 + E_{\a_0}^1
\lab{gz}
\ee
with $a=1,2,\ldots p-1$ and $g$ given in \rf{gd} ($g(z)^p=z \one$), we can
then write the generators \rf{h0}-\rf{step} as
\br
\chh^{0}_{k,pn} &\equiv& z^n \( \om_k-1\) D^k + \d_{n,0} C \one
\nonu\\
\chh^{l+1}_{k,pn+k} &\equiv& z^n \om_{kl} D^{kl} (g(z))^k \sim z^n \( D^{l}
g(z)\)^k
\lab{basis}
\er
with $l=0,1,\ldots p-1$ and $k=1,2, \ldots p-1$.

The orthogonality of the subspaces \rf{kmsubspaces} is shown by using the same
arguments as the ones given above in the finite dimensional case, but now
using the  bilinear form
\be
\langle A(z) \mid B(z) \rangle \equiv
\oint {dz \o 2 \pi i z} \Tr (A(z)  B(z))
\ee

Using \rf{bracket} one obtains
\br
\lb \chh^{0}_{j,pm} \, ,\, \chh^{0}_{k,pn}\rb &=& C\,
\,(\om_{j}-1)(\om_{p-j}-1)\,  p m\, \d_{n+m,0} \d_{k,p-j}
\lab{heis1}\\
\lb \chh^{l+1}_{j,pm+j} \, ,\, \chh^{l+1}_{k,pn+k}\rb &=& C\, \om_l^{k+j+kj} \,
(pm+j)\, \d_{m+n+1,0} \d_{k,p-j}
\lab{heis2}
\er
which constitute a set of $p+1$ orthogonal Heisenberg subalgebras.

We now comment on a connection between the decomposition above and a special
basis for the affine Kac-Moody algebras. Such a basis \cite{kklw,kac} is a
generalization of the basis constructed by Kostant \cite{kostant} for finite
dimesional simple Lie algebras. Consider the principal gradation of an affine
Kac-Moody algebra $\cgh$
\be
\cgh = \oplus \, \cgh_n
\ee
with $n\in \IZ$ and
\be
\lb Q_{\mbox{\rm pr.}} \, , \, \cgh_n \rb = n \,\cgh_n
\ee
and $Q_{\mbox{\rm pr.}}$ being the grading operator \rf{gradop} corresponding
to  all $s_i$'s equal to $1$. Let $\chh \subset \cgh$ be a principal Heisenberg
subalgebra with generators $E_M \in \cgh_M$, where $M$ are the exponents of
$\cgh$, and satisfying
\be
\lb E_M \, , \, E_N \rb = {C \o h} \Tr (E_M E_N ) M \d_{M+N,0}
\ee
These exponents have a period equal to the Coxeter number $h$ of the finite
simple Lie algebra $\cg$ (of which $\cgh$ is the affinization of), i.e. they
have the form $M=m_a + nh$ where $n$ is an integer and $m_a$, $a=1,2,\ldots r$
are the exponents of $\cg$ \cite{kac}. In particular, unity is always an
exponent and it follows that $E_1$ is a linear combination of the simple root
step operators
$E_{\a_a}^0$ and $E_{\a_0}^1$, $a=1,2, \ldots r$. The complement $\cf$ of
$\chh$ in $\cgh$ is such that the dimension of $\cf_m \subset \cgh_m$ for $m
\ne 0$ is equal to the rank of $\lie$ ($\equiv r$). The subspace $\cf_0$ has
dimension $r+2$ and is generated by $C$, $D$ and $H_a^0$, $a=1,2, \ldots, r$.
The complement $\cf$ ad-diagonalize $\chh$ and except for the extra
two generators of $\cf_0$, the basis, $F^a_m$, of
$\cf_m$  can be chosen such that
\be
\lb E_M \, , \, F^a_n \rb \sim F^a_{M+n} \lab{ef}
\ee

We point out that for the affine algebras $\hat{sl}(p, \IC )$ with $p$ prime,
the complement $\cf$ is in fact an algebraic sum of $p$ Heisenberg
subalgebras, with $p-1$ of them being principal and $1$ being homogeneous. In
addition, the role played by $\chh$ can be replaced by any one of these $p-1$
principal Heisenberg subalgebras, with $\chh$ being now part of the new
complement $\cf$.

To establish this fact we show that by choosing one the principal Heisenberg
subalgebras $\chh_{l+1}$ in \rf{kmsubspaces} to play the role of $\chh$ as
above, then the basis elements of the remaining Heisenberg subalgebras
ad-diagonalize $\chh$.

Using \rf{bracket} and \rf{basis} one gets
\be
\lb \chh_{j,pm+j}^{l+1} \, , \, \chh_{k,pn}^0 \rb = (\om_{kj}-1)(1-\om_{p-k})
\chh_{j,p(m+n)+j}^{l^{\pr}+1}
\lab{comp1}
\ee
where $l^{\pr}$ is the integer such that $(j(l^{\pr}-l)-k)$ is a multiple of
$p$. Notice that $l^{\pr}$ and $l$ are never equal since  $k$ is not allowed to
be a multiple of $p$ (see \rf{basis}). In addition one gets ($l\neq l^{\pr}$)
\br
\lb \chh_{j,pm+j}^{l+1} \, , \, \chh_{k,pn+k}^{l^{\pr}+1} \rb =
(\om_{l^{\pr}}^{jk}-\om_l^{jk})\chh_{k+j,p(m+n)+k+j}^{l^{\pr\pr}+1}
\qquad \mbox{\rm if $k+j\neq p$}
\lab{comp2}
\er
where $l^{\pr\pr}$ is the integer such that $k(l^{\pr\pr}-l^{\pr}) +
j(l^{\pr\pr}-l)$ is a multiple of $p$, and (again $l\neq l^{\pr}$)
\br
\lb \chh_{j,pm+j}^{l+1} \, , \, \chh_{k,pn+k}^{l^{\pr}+1} \rb =
{\om_{j(l-l^{\pr})}(\om_{l^{\pr}}^{-j^2}-\om_l^{-j^2})\o
(\om_{j(l-l^{\pr})}-1)} \chh_{j(l-l^{\pr}),p(m+n+1)}^0
\qquad \mbox{\rm if $k+j= p$}
\lab{comp3}
\er
Notice that in \rf{comp2} $l^{\pr\pr}$ can not be equal to $l$ or $l^{\pr}$,
since if it is equal, let us say, to $l$ then either $k$ or
$(l^{\pr\pr}-l^{\pr})$ would have to have $p$ in its expansion as a product of
primes, and that is not allowed. We point out that the term $\d_{n,0}\, C\,
\one$ in the expression of $\chh_{k,pn}^0$ given by \rf{basis}, was not
important in obtaining
\rf{heis1} or the orthogonality property. However it was crucial in
establishing \rf{comp3}.

Therefore the elements of the $p+1$ Heisenberg subalgebras \rf{kmsubspaces}
ad-diagonalize any one of the $p$ principal Heisenberg subalgebras
$\chh_{l+1}$, $l=0,1, \ldots p-1$. They therefore constitute a basis for the
complement $\cf$ discussed above. Notice however that the basis elements
$F^a_n$ for a fixed $a$ are not all in the same Heisenberg subalgebra.
In addition, the results of brackets between elements of two given Heisenberg
subalgebras  do not belong necessarily to a fixed third Heisenberg subalgebra,
in other words the decomposition is not multiplicative orthogonal in
the sense of ref. \cite{KKU82}.

We end this section by making a comment on a fine grading of $A_2$ \ct{PZ81}.
It is a $\IZ_3\times \IZ_3$ gradation where the generators have the following
gradings:
$D=(1,0)$, $D^2=(2,0)$, $g=(0,2)$, $g^2=(0,1)$, $Dg=(1,2)$, $(Dg)^2=(2,1)$,
$D^2g=(2,2)$,and $(D^2g)^2=(1,1)$, with $D$ and $g$ being the $3\times 3$
matrices defined in \rf{gd}. The addition of the gradings is given by the sum
mod $3$. The operator with grade $(0,0)$ is the identity matrix, which is not
part of the algebra $A_2$ (one has to extend to $gl(3)$ to incorporate it).
Since this gradation respects the  decomposotion \rf{decom} for $A_2$ one can
elevate it to the corresponding affine algebra. One can just attach an integer
index $n$ to the generators (eigenvalue of the derivation $d$, see \rf{km}) and
get a gradation $\IZ_3\times \IZ_3 \times \IZ$ of the Kac-Moody algebra
$A_2^{(1)}$. Such a gradation is not of the type \rf{gradop}.  However, the
decomposition in terms of Heisenberg subalgebras
\rf{decomkm} we get in this case is the same as the one we would get with the
homogeneous gradation.

\section{The case $A_{p^m-1}$ with $p$ prime and $m>1$}

In refs. \cite{KKU81}, \cite{KKU811} a construction of the orthogonal
decompositions for the algebras $A_{p^m-1}$ with $p$ prime and $m>1$ was also
performed.  In the case $m=1$ (see previous section) the basic idea was that
matrices made of powers of a given matrix $M$ commute among themselves. For the
case $m>1$ one uses the fact that, given the matrices $M_i$, $i=1,2,\ldots ,m$,
then the quantities $M_1^{n_1}\otimes  M_2^{n_2}\otimes \ldots \otimes
M_m^{n_m}$, commute among themselves under the usual commutator in the
tensor product
\br
 \lb A_1 \otimes  \ldots \otimes A_m \, ,\, B_1 \otimes  \ldots
\otimes B_m \rb \equiv
 A_1 B_1 \otimes \ldots \otimes A_m B_m -
B_1 A_1 \otimes \ldots \otimes B_m A_m
\lab{tensorcom}
\er
Therefore for the complex Lie algebra $A_{p^m-1}$ with $p$ prime and $m>1$ we
introduce the following subspaces
\br
\ch_0 &\equiv& \{ D^{k_1}\otimes D^{k_2}\otimes \ldots \otimes D^{k_m}\, \mid
\,  k_i=0,1,2,\ldots ,p-1  \} \nonu\\
\ch_{\bf L}&\equiv& \{ (D^{l_1}g)^{k_1}\otimes (D^{l_2}g)^{k_2}\otimes \ldots
\otimes (D^{l_m}g)^{k_m}\, \mid \, k_i=0,1,2,\ldots ,p-1 \}
\lab{subpm}
\er
where $D$ and $g$ are the $p\times p$ matrices defined in \rf{gd}, and ${\bf L}
\equiv (l_1,l_2,\ldots l_{m})$, with $l_i=0,1,2,\ldots ,p-1$ . In addition the
$k_i$'s are not all allowed to have the value zero.

Notice that there are $p^m$ possible vectors ${\bf L}$ and that the dimension
of $\ch_0$ and $\ch_{\bf a}$ is $p^m-1$. Therefore we have $(p^m-1)(p^m+1)$
generators in those subspaces, which is the  dimension of $A_{p^m-1}$. Each
one the above subspaces is obviously abelian. For $p$ being a prime number
those generators are linearly independent and constitute a basis for
$A_{p^m-1}$. The proof for that and the orthogonality property is very similar
for the case $m=1$, given in the previous section, and for more details we
refer
to \cite{KKU811}.

In order to extend the above decomposition to the corresponding affine
Kac-Moody algebra we have to find a gradation which respects the decomposition
in terms of \rf{subpm}. We now show that the only two allowed gradations are
given by
\br
{\bf s}_{\rm hom} &=& \( 1,0,0, \ldots ,0\) , \qquad \mbox{\rm homogeneous
gradation};
\nonu\\
{\bf s}_{\rm p} &=& \( 1,\underbrace{0, \ldots ,0,1}_{p{\rm\; entries}},
\underbrace{0, \ldots ,0,1}_{p{\rm\; entries}},\ldots\ldots
\underbrace{0, \ldots ,0,1}_{p{\rm\; entries}},
\underbrace{0, \ldots ,0}_{p-1{\rm\; entries}} \),
\lab{allgrad}
\er
where we have denoted the integers $s_a$ in \rf{gradop} as the components of
the vector ${\bf s} \equiv \( s_0, s_1, \ldots ,s_{p^m-1}\)$.

To show that, it is better to express the elements of the subspaces
\rf{subpm} in terms of step operators. The elements of $\ch_0$ are diagonal and
therefore are linear combinations of the Cartan subalgebra generators only.
Therefore, under \rf{gradop}, they will be sensitive to the $d$ term only.
Consequently, when affinizing the algebra what one has to do is to make sure
all Cartan subalgebras generators appearing in the expansion of the
affinized version of a given element of $\ch_0$ have the same $d$-eigenvalue.
Therefore $\ch_0$ does not impose any restrictions on the possible gradations.

As for the elements of $\ch_{\bf L}$ we can write, using \rf{prop},
$(D^{l_1}g)^{k_1}\otimes (D^{l_2}g)^{k_2}\otimes \ldots \otimes
(D^{l_m}g)^{k_m}
\sim (D^{l_1 k_1}\otimes D^{l_2 k_2}\otimes \ldots \otimes D^{l_m k_m})
(g^{k_1}\otimes  g^{k_2}\otimes \ldots \otimes g^{k_m})$. When expanding such
an element in the Chevalley basis, the step operator content is determined by
the
term $(g^{k_1}\otimes  g^{k_2}\otimes \ldots \otimes g^{k_m})$, since the other
one just dictates which linear combination of these step operators it
corresponds to. Therefore in order to determine the gradings respecting the
decomposition \rf{subpm}, one has to consider only the $g$ part of those
elements only.

Consider the quantity
$\one \otimes \one \otimes \ldots \otimes\one \otimes g$, which, written in
terms of $p^m\times p^m$ matrices,  has the form
\br
\one \otimes \one \otimes\ldots \otimes\one \otimes g = \(
\begin{array}{ccccc}
g&0&0&\cdots&0\\
0&g&0&\cdots&0\\
\vdots&\vdots&\vdots&\vdots&\vdots\\
0&0&0&\cdots&g
\end{array}\)
\lab{11g}
\er
where $g$ is the matrix given in \rf{gd}. Using the matrices for step
operators in the defining representation of $A_{p^m-1}$, given in \rf{defrep}
(excluding the $z$ terms), one concludes that the above element contains in
its expansion all the step operators for simple roots $E_{\a_a}$, except
those corresponding to $a=k p$, with $k=1,2,\ldots p^{m-1}-1$. When affinizing
such an element we will need to find integers $n_a$ and $n_b$ such that
$E_{\a_a}^{n_a}$ and $E_{\a_b}^{n_b}$ have the same grade under the
grading operator \rf{gradop}. Using arguments similar to those of the previous
section (see paragraph before eq. \rf{h0}) one then concludes that the gradings
respecting the decomposition \rf{subpm} are such that $s_a = s_b$ for all $a$
and $b$ which are not multiples of $p$.

{}From \rf{11g}, \rf{gd} and \rf{defrep} one observes that $\one \otimes \one
\otimes \ldots \otimes\one \otimes g$ also contains in its expansion step
operators corresponding to the lowest root of each block, namely
$E_{-\a_{jp+1}-\a_{jp+2}-\ldots -\a_{jp+p-1}}$, $j=0,1,\ldots p^{m-1}-1$.
Therefore we have to find integers $n$ and ${\bar n}$ such that
$E_{\a_{jp+s}}^n$ ($s=1,2,\ldots p-1$) and $E_{-\a_{jp+1}-\a_{jp+2}-\ldots
-\a_{jp+p-1}}^{{\bar n}}$  have the same grade under the grading operator
\rf{gradop}. Therefore we need $s_{jp+s} + N n = - \sum_{q=1}^{p-1} s_{jp+q} +
N {\bar n}$, or ${\bar n} = n + {p s_{jp+s}\o N}$, where we have used the fact
that, according to the reasoning above, all $s_{jp+q}$'s ($q=1,2,\ldots p-1$)
are equal. But, by the same reasons, we have $N= s_0 + p^{m-1}(p-1) s_{jp+s} +
\sum_{k=1}^{p^{m-1}-1} s_{kp}$. Therefore, since all $s_i$'s are non negative,
it follows that ${p s_{jp+s}\o N}$ can not be an integer unless
\be
s_{jp+s}=0, \qquad \mbox{\rm for $j=0,1,\ldots p^{m-1}-1$ and $s=1,2,\ldots
p-1$}
\lab{first}
\ee
The case $p=2$ with $m=2$ may lead to other possibilities, but one can check
that due to the symmetries of the $A_{3}$ Dynkin diagram they do not lead to
gradings inequivalent to those given in \rf{allgrad}.

Consider now the quantity
\br
\one \otimes \one \otimes\ldots \otimes\one \otimes g \otimes g = \(
\begin{array}{ccccc}
0&g&0&\cdots&0\\
0&0&g&\cdots&0\\
\vdots&\vdots&\vdots&\vdots&\vdots\\
0&0&0&\cdots&g\\
g&0&0&\cdots&0
\end{array}\)
\lab{gg}
\er
where again $g$ is the matrix given in \rf{gd}, but now to simplify we take
all $x_i$'s  equal to unity. Using \rf{defrep} one observes that such an
element
contains in its expansion all step operators for simple roots not appering in
\rf{11g}, namely $E_{\a_{kp}}$ with $k=1,2,\ldots p^{m-1}-1$. Therefore we need
to find integers $n_k$ and $n_{\bar k}$ such that $E_{\a_{kp}}^{n^k}$ and
$E_{\a_{{\bar k}p}}^{n_{\bar k}}$ have the same grade under the grading
operator \rf{gradop}. Using the same arguments as above one concludes that all
$s_{kp}$'s ($k=1,2,\ldots p^{m-1}-1$) must be equal.

One observes that the step operator for the  lowest root of $A_{p^m-1}$ also
appears in \rf{gg}. Therefore we need integers $n_k$ and $n_{\psi}$ such that
$E_{\a_{kp}}^{n_k}$ and $E_{-\psi}^{n_{\psi}}$ ($\psi = \sum_{a=1}^{p^m-1}
\a_{a}$) have the same grade under the grading operator \rf{gradop}.  So, we
need $n_{\psi} = n_k + {p^{m-1} s_{kp}\o N}$, where $N=s_0 +
(p^{m-1}-1)s_{kp}$, and where we have used \rf{first} and the fact that all
$s_{kp}$'s are equal.One then concludes there are only two ways for the
ratio ${p^{m-1} s_{kp}\o N}$ to be an integer, namely
\be
s_{kp}=0 \qquad \mbox{\rm for $k=1,2,\ldots p^{m-1}-1$}
\ee
or
\be
s_{kp}=s_0 \qquad \mbox{\rm for $k=1,2,\ldots p^{m-1}-1$}
\ee
Again the case $p=2$ with $m=2$ leads to other possibilities but
they do not give inequivalent gradations. But those possibilities correspond
exactly to the two  gradations given in \rf{allgrad}. One can check that by
considering the other elements of $\ch_{\bf L}$ one does not get any new
restrictions on the allowed gradations, showing that \rf{allgrad} are the only
two possibilities.

Therefore, if we use the homogeneous gradation ${\bf s}_{\rm hom}$ of
\rf{allgrad}, the subspaces \rf{subpm} will give rise to $p^m+1$ homogeneous
Heisenberg subalgebras. On the other hand if we use the gradation ${\bf s}_{\rm
p}$, the subspace $\ch_0$ will lead to a homogeneous Heisenberg subalgebra,
and the subspaces $\ch_{{\bf L}}$ to $p^m$ Heisenberg subalgebras associated
to the grading ${\bf s}_{\rm p}$.

The process of writing the elements of the subspaces \rf{subpm} in the
Chevalley basis is quite involved. However, there are some general patterns
which are useful. As we commented in the paragraph before \rf{11g}, the step
operator content of those elements is determined by the terms $g^{k_1}\otimes
g^{k_2}\otimes \ldots \otimes g^{k_m}$. These, when written in terms of
$p^m\times p^m$ matrices, will be made of blocks which are powers of  the
$p\times p$ matrix $g$ introduced in \rf{gd}. Using \rf{defrep} one can
check that such a structure implies the step operators appearing in the
expasion of a given element of the subspaces $\ch_{{\bf L}}$ in
\rf{subpm} can be arranged in sets of $p$ step operators such that the
corresponding roots add up  to zero. Below we discuss the cases
$A_{p^m-1}$ with $m=2$ and $p=2,3$.

\subsection{The case $A_{3}$}

We denote the three simple roots of $A_{3}$ as $\a_a$, $a=1,2,3$, such that
$2\a_1 \cdot \a_2/ \a_2^2 = 2\a_3 \cdot \a_2/ \a_2^2=-1$ and $\a_1 \cdot
\a_3 =0$. We consider here the gradation ${\bf s}_{\rm p}$ in \rf{allgrad},
i.e.
\be
{\bf s}_{\rm p}=(1,0,1,0)
\ee
The corresponding results for the homogeneous gradation can be easily obtained
from those for  ${\bf s}_{\rm p}$. The generators for the affine subspace
$\ch_0$ in \rf{subpm} are obviously $H_a^n$, $a=1,2,3$.
The commutativity of the subspaces $\ch_{{\bf L}}$ and their pairwise
orthogonality condition impose very rigid constraints on the
allowed combinations of step operators. Now it is a quite simple algebraic task
to convince oneself that the remaining $4$ affine subspaces $\ch_{{\bf L}}$ are
\begin{eqnarray}
\chh_1 &=&
\{ (E_{\alpha_1}^n+E_{-\alpha_1}^n)+
(E_{\alpha_3}^n+E_{-\alpha_3}^n),\;
(E_{\alpha_1+\alpha_2}^n+E_{-\alpha_1-\alpha_2}^{n+1})+
(E_{\alpha_2+\alpha_3}^n+E_{-\alpha_2-\alpha_3}^{n+1}),\;\nonu\\
& &(E_{\alpha_1+\alpha_2+\alpha_3}^n+E_{-\alpha_1-\alpha_2-\alpha_3}^{n+1})+
(E_{\alpha_2}^n+E_{-\alpha_2}^{n+1})\} ;\nonumber \\
\chh_2 &=&
\{ (E_{\alpha_1}^n+E_{-\alpha_1}^n)-
(E_{\alpha_3}^n+E_{-\alpha_3}^n),\;
(E_{\alpha_1+\alpha_2}^n-E_{-\alpha_1-\alpha_2}^{n+1})+
(E_{\alpha_2+\alpha_3}^n-E_{-\alpha_2-\alpha_3}^{n+1}),\;\nonu\\
& &(E_{\alpha_1+\alpha_2+\alpha_3}^n-E_{-\alpha_1-\alpha_2-\alpha_3}^{n+1})+
(E_{\alpha_2}^n-E_{-\alpha_2}^{n+1})\} ;\nonumber \\
\chh_3 &=&
\{ (E_{\alpha_1}^n-E_{-\alpha_1}^n)+
(E_{\alpha_3}^n-E_{-\alpha_3}^n),\;
(E_{\alpha_1+\alpha_2}^n+E_{-\alpha_1-\alpha_2}^{n+1})-
(E_{\alpha_2+\alpha_3}^n+E_{-\alpha_2-\alpha_3}^{n+1}),\;\nonu\\
& &(E_{\alpha_1+\alpha_2+\alpha_3}^n-E_{-\alpha_1-\alpha_2-\alpha_3}^{n+1})-
(E_{\alpha_2}^n-E_{-\alpha_2}^{n+1})\} ;\nonumber \\
\chh_4 &=&
\{ (E_{\alpha_1}^n-E_{-\alpha_1}^n)-
(E_{\alpha_3}^n-E_{-\alpha_3}^n),\;
(E_{\alpha_1+\alpha_2}^n-E_{-\alpha_1-\alpha_2}^{n+1})-
(E_{\alpha_2+\alpha_3}^n-E_{-\alpha_2-\alpha_3}^{n+1}),\;\nonu\\
& &(E_{\alpha_1+\alpha_2+\alpha_3}^n+E_{-\alpha_1-\alpha_2-\alpha_3}^{n+1})-
(E_{\alpha_2}^n+E_{-\alpha_2}^{n+1})\} .
\label{A3}
\end{eqnarray}

The corresponding grading operator is (see \rf{gradop})
\be
Q_{{\bf s}_{\rm p}}= {2\lambda_2\o \a_2^2}\cdot H + 2 d
\ee
and the generators are split into odd and even eigenvalues, i.e. introducing
the subspaces
\begin{eqnarray}
\cgh_{2n} &\equiv &
\{ E_{\alpha_1}^n, E_{-\alpha_1}^n; E_{\alpha_3}^n, E_{-\alpha_3}^n;
H_a^n \, \, (a=1,2,3)\} ,
\label{A3'}\\
\cgh_{2n+1} &\equiv &
\{ E_{\alpha_1+\alpha_2}^n, E_{-\alpha_1-\alpha_2}^{n+1};
E_{\alpha_2+\alpha_3}^n, E_{-\alpha_2-\alpha_3}^{n+1} ,
E_{\alpha_1+\alpha_2+\alpha_3}^n, E_{-\alpha_1-\alpha_2-\alpha_3}^{n+1};
E_{\alpha_2}^n, E_{-\alpha_2}^{n+1}\} ,
\nonumber
\end{eqnarray}
one has
\br
\lb Q_{{\bf s}_{\rm p}} \, , \,  \cgh_{2n} \rb
&=& 2n\, \cgh_{2n} \nonu\\
\lb Q_{{\bf s}_{\rm p}} \, , \, \cgh_{2n+1} \rb
&=& (2n+1)\, \cgh_{2n+1}
\er

\subsection{The case $A_{8}$}

As usual we denote the simple roots of $A_{8}$ as $\a_a$, $a=1,2,\ldots,8$,
such that $\a_1$ and $\a_8$ correspond to the end points of the Dynkin diagram
of $A_{8}$. Again we consider the gradation ${\bf s}_{\rm p}$ in
\rf{allgrad}, i.e.
\be
{\bf s}_{\rm p}=(1,0,0,1,0,0,1,0,0)
\ee
The corresponding grading operator \rf{gradop} is then
\be
Q_{{\bf s}_{\rm p}}= {2\lambda_3\o \a_3^2}\cdot H +
{2\lambda_6\o \a_6^2}\cdot H + 3 d
\lab{grada8}
\ee

The affine subspace $\ch_0$ in \rf{subpm} is again generated by
$H_a^n$, $a=1,2,\ldots ,8$, and the generators  of the $9$ affine subspaces
$\ch_{{\bf L}}$ are linear combinations of the following sets of step operators
\br
\cgh^{(1)}_{3n}&=&
\{ E_{\a_1}^n, E_{\a_2}^n, E_{-\a_1-\a_2}^n; E_{\a_4}^n, E_{\a_5}^n,
E_{-\a_4-\a_5}^n;  E_{\a_7}^n, E_{\a_8}^n, E_{-\a_7-\a_8}^n\} ,
\nonu\\
\cgh^{(2)}_{3n+1}&=&
\{ E_{\a_2+\a_3}^n, E_{\a_4+\a_5+\a_6+\a_7+\a_8}^n,
E_{-\a_2-\a_3-\a_4-\a_5-\a_6-\a_7-\a_8}^{n+1};
\nonu\\
& &
E_{\a_3+\a_4}^n, E_{\a_5+\a_6}^n, E_{-\a_3-\a_4-\a_5-\a_6}^{n+1};
\nonu\\
& &
E_{\a_6+\a_7}^n,
E_{\a_1+\a_2+\a_3+\a_4+\a_5}^n, E_{-\a_1-\a_2-\a_3-\a_4-\a_5-\a_6-
\a_7}^{n+1}\} ,
\nonu\\
\cgh^{(3)}_{3n+1}&=&
\{ E_{\a_1+\a_2+\a_3}^{n}, E_{\a_4+\a_5+\a_6}^{n},
E_{-\a_1-\a_2-\a_3-\a_4-\a_5-\a_6}^{n+1};
\nonu\\
& &
E_{\a_2+\a_3+\a_4}^{n},
E_{\a_5+\a_6+\a_7}^{n}, E_{-\a_2-\a_3-\a_4-\a_5-\a_6-\a_7}^{n+1};
\nonu\\
& &
E_{\a_3+\a_4+\a_5}^{n}, E_{\a_6+\a_7+\a_8}^{n},
E_{-\a_3-\a_4-\a_5-\a_6-\a_7-\a_8}^{n+1} \} ,
\nonu\\
\cgh^{(4)}_{3n+1}&=&
\{ E_{\a_1+\a_2+\a_3+\a_4}^{n}, E_{\a_5+\a_6+\a_7+\a_8}^{n},
E_{-\a_1-\a_2-\a_3-\a_4-\a_5-\a_6-\a_7-\a_8}^{n+1};
\nonu\\
& &
E_{\a_2+\a_3+\a_4+\a_5}^{n}, E_{\a_6}^{n}, E_{-\a_2-\a_3-\a_4-\a_5-\a_6}^{n+1};
\nonu\\
& &
E_{\a_3}^{n}+E_{\a_4+\a_5+\a_6+\a_7}^{n}, E_{-\a_3-\a_4-\a_5-\a_6-
\a_7}^{n+1}\} ,
\er
and $4$ other sets of elements which are conjugated to these by
$E_{\pm\alpha}^n \ra E_{\mp \alpha}^{-n}$. The lower indices on $\cgh$'s are
the eigenvalues w.r.t. the grading operator \rf{grada8}.
Notice that each of the $8$ subspaces $\cgh$'s contains $9$ elements. The
generators of the $9$ Heisenberg subalgebras $\ch_{{\bf L}}$ will be linear
combinations of the $9$ elements of each one of those subspaces. And  each one
of the $8$ elements of a given Heisenberg subalgebra comes from a
different subspace $\cgh^{(i)}$.

\section{The case of $G_2$}

The authors of ref. \cite{KKU82} considered also what they called {\em a root
orthogonal decomposition} ${\cal R}{\cal O}{\cal D}$. In this case the algebra
is decomposed as in \rf{decom} with the subspace $\ch_0$ being the usual Cartan
subalgebra and the remaining ones being generated by operators of the form
$E_{\a}\pm E_{-\a}$, with $\a$ being a positive root of $\cg$. Clearly any two
roots, $\a$ and $\b$ appearing in a given subspace $\ch_i$, must be such that
$\pm\a\pm \b$ are not  roots.

The complex Lie algebra $G_2$ was shown, in ref. \cite{KKU82}, to
possess ${\cal R}{\cal O}{\cal D}$. Denote the roots of $G_2$ as $\a_1$,
$\a_2$, $\a_1+\a_2$, $2\a_1+\a_2$, $3\a_1+\a_2$ and $3\a_1+2\a_2$, with $\a_1$
and
$\a_2$ being the simple roots. The seven subspaces in the ${\cal R}{\cal
O}{\cal
D}$ are
\br
\ch_0 &=& \{ H_1 \, , \, H_2 \}\nonu\\
\ch_1^{\pm} &=& \{ E_{\a_1}\pm E_{-\a_1} \, , \, E_{3\a_1+2\a_2}\pm
E_{-3\a_1-2\a_2} \}\nonu\\
\ch_2^{\pm} &=& \{ E_{\a_1+\a_2}\pm E_{-\a_1-\a_2} \, , \, E_{3\a_1+\a_2}\pm
E_{-3\a_1-\a_2} \}\nonu\\
\ch_3^{\pm} &=& \{ E_{\a_2}\pm E_{-\a_2} \, , \, E_{2\a_1+\a_2}\pm
E_{-2\a_1-\a_2} \}
\er
where $H_a$, $a=1,2$, and $E_{\pm \a}$ are the Cartan subalgebra generators and
step operators respectivally in the Chevalley basis. Those subspaces are
clearly abelian and orthogonal. In addition, that decomposition is also {\em
multiplicative orthogonal} in the sense that for any $i$ and $j$ there exists a
$k$ such that $\lb \ch_i \, , \, \ch_j \rb \subset \ch_k$.

In order to extend this decomposition to the affine Kac-Moody algebra ${\hat
G_2}$ we have to find the gradings that respect the decompostion. Since the
subspaces are made of linear combinations of positive and negative step
operators associated to each root, the grading has to be such that for every
positive root $\a$ there are integers $n_{\pm}$ such that $E_{\a}^{n_+}$ and
$E_{-\a}^{n_-}$ have the same grades. From \rf{gradop} and \rf{grad} we see
that for a root $\a = m_1\a_1 + m_2 \a_2$ we need $(n_- - n_+) = 2(m_1 s_1   +
m_2 s_2)/N \equiv \mbox{\rm integer}$, where $N=s_0+3s_1+2s_2$ (since $\psi =
3\a_1 + 2\a_2$). For $\a \equiv \a_1$ we observe that this is impossible if
$s_1\neq 0$. Now taking $s_1=0$, we get for $\a \equiv m_1 \a_1 + \a_2$ that
$(n_- - n_+)$ is  integer only if $s_0=0$ or then if $s_2=0$. For $\a \equiv 3
\a_1 + 2 \a_2$ these two possibilities are also allowed. So there are two
gradations which respect the decomposition above, namely,
\br
(s_0,s_1,s_2)&=&(1,0,0) \qquad\mbox{\rm  homogeneous gradation} \nonu\\
(s_0,s_1,s_2)&=&(0,0,1)
\er
which are not equivalent, since there are no symmetries of the Dynkin diagram
relating them. We shall refer to the second gradation above as $Q_2$
gradation.

We will consider here the gradation $Q_2$, since the same results can be easily
translated to the homogeneous gradation. We define the generators
\br
\chh_{0,1}^{(2n)} &\equiv& (2 H_1^n+3 H_2^n)/\sqrt{3}  \nonu\\
\chh_{0,2}^{(2n)} &\equiv& H_2^n  \nonu\\
\chh_{1,1}^{\pm,(2n)} &\equiv& (E_{\a_1}^n \pm E_{-\a_1}^n)/i^{\sigma}\sqrt{3}
\nonu\\
\chh_{1,2}^{\pm,(2n)} &\equiv& (E_{3\a_1+2\a_2}^{n-1} \pm
E_{-3\a_1-2\a_2}^{n+1})/i^{\sigma}
\nonu\\
\chh_{2,1}^{\pm,(2n+1)} &\equiv& (E_{\a_1+\a_2}^{n} \pm
E_{-\a_1-\a_2}^{n+1})/i^{\sigma}\sqrt{3}
\nonu\\
\chh_{2,2}^{\pm,(2n+1)} &\equiv& (E_{3\a_1+\a_2}^{n} \pm
E_{-3\a_1-\a_2}^{n+1})/i^{\sigma}
\nonu\\
\chh_{3,1}^{\pm,(2n+1)} &\equiv& (E_{\a_2}^{n} \pm E_{-\a_2}^{n+1})/i^{\sigma}
\nonu\\
\chh_{3,2}^{\pm,(2n+1)} &\equiv& (E_{2\a_1+\a_2}^{n} \pm
E_{-2\a_1-\a_2}^{n+1})/i^{\sigma}\sqrt{3}
\lab{basisg2}
\er
where $\sigma = 0$ for the $+$ sign and $\sigma = 1$ for the $-$ sign. These
generators  are all eigenvectors of the grading operator (see \rf{gradop})
\be
Q_2 \equiv \lambda^v_2 \cdot H + 2 d
\ee
and the eigenvalues are the numbers between parentheses in the upper indices in
\rf{basisg2}.

The affine Kac-Moody algebra ${\hat G_2}$ can then be decomposed
into
\be
{\hat G_2}=\chh_0\oplus \chh_1^+\oplus \chh_1^-\oplus
\chh_2^+\oplus\chh_2^-\oplus\chh_3^+\oplus\chh_3^-
\ee
with
\br
\chh_0 &\equiv& \{ \chh_{0,a}^{(2n)} \, , \, a=1,2 \, , \, n \in \IZ\} \nonu\\
\chh_1^{\pm}&\equiv& \{ \chh_{1,a}^{\pm,(2n)} \, , \, a=1,2\, , \, n \in \IZ\}
\nonu\\
\chh_2^{\pm}&\equiv& \{ \chh_{2,a}^{\pm,(2n+1)} \, , \, a=1,2\, , \, n \in
\IZ\}
\nonu\\
\chh_3^{\pm}&\equiv& \{ \chh_{3,a}^{\pm,(2n+1)} \, , \, a=1,2\, , \, n \in
\IZ\}
\er

Using the bilinear form
\br
\Tr ( H_a^m \, H_b^n) = \eta_{ab} \d_{m+n,0} \qquad
\Tr ( E_{\a}^m \, E_{\b}^n) = {2\o \a^2} \d_{\a +\b ,0}\d_{m+n,0}
\er
where
\br
\eta_{ab} \equiv {2\o \a_a^2} K_{ab} =
\(
\begin{array}{cc}
6&-3\\
-3&2
\end{array}\)
\er
and where we have normalized the roots as $\a_1^2 = 2/3$ and $\a_2^2=2$, one
can check that the above subspaces are orthogonal.

Using \rf{km} one can check that
\br
\lb \chh_{0,a}^{(2m)} \, , \, \chh_{0,b}^{(2n)} \rb &=& C\, 2m\,
\d_{a,b}\,\d_{m+n,0}\nonu\\
\lb \chh_{1,a}^{\epsilon ,(2m)} \, , \, \chh_{1,b}^{\epsilon ,(2n)} \rb &=& C\,
2m\, \d_{a,b}\,\d_{m+n,0}\nonu\\
\lb \chh_{2,a}^{\epsilon ,(2m+1)} \, , \, \chh_{2,b}^{\epsilon ,(2n+1)} \rb &=&
C\, (2m+1)\, \d_{a,b}\,\d_{m+n+1,0}\nonu\\
\lb \chh_{3,a}^{\epsilon ,(2m+1)} \, , \, \chh_{3,b}^{\epsilon ,(2n+1)} \rb &=&
C\, (2m+1)\, \d_{a,b}\,\d_{m+n+1,0}
\er
where $\epsilon \equiv \pm$ and $a,b=1,2$. Therefore we have seven orthogonal
Heisenberg  subalgebras.

In addition, using the relations \rf{km} with the fact that $\epsilon (\a \, ,
\, \b ) = - \epsilon (-\a \, ,\, -\b )$, one can check that
\br
\lb \chh_i^{+} \, , \, \chh_i^{-} \rb &\subset& \chh_0  \nonu\\
\lb \chh_0 \, , \, \chh_i^{\pm} \rb &\subset& \chh_i^{\mp}  \nonu\\
\lb \chh_i^{\epsilon_1} \, , \, \chh_j^{\epsilon_2} \rb &\subset&
\chh_k^{-\epsilon_1\epsilon_2}
\er
where $i,j,k=1,2,3$ and in the last relation $i\neq j\neq k$; $\epsilon_1 ,
\epsilon_2 = \pm$.

\section{Conclusion}

In the present paper we studied the decomposability problem
of the affine Lie algebras into the algebraic sum of pairwise
orthogonal Heisenberg subalgebras. In particular, the
Kostrikin--Kostrikin--Ufnarovskii theorem about the orthogonal
decomposition of some complex simple Lie algebras can be extended
to the case of the corresponding affine Lie algebras endowed with
the homogeneous gradation. However, the most interesting situation
arises when other gradations also respect such a rigid decomposition,
and we investigated this possibility in detail by the examples of the Lie
algebras of type $A_r^{(1)}$ and $G_2^{(1)}$. Unfortunately, as in the
finite--dimensional case, the consideration itself crucially depends
on an explicit realization of the algebras in question, and is performed
separately for various types of them.
So, it is very interesting to find an uniform formulation for all
algebras which admit the orthogonal (and multiplicative) decomposition.

An interesting issue stimulated by the existence of such decompositions
 concerns the representation
theory of these special affine Kac-Moody algebras. As  is well known, the
Heisenberg subalgebra is the algebra of oscillators in quantum field theory.
So, one could conjecture that representations for these algebras could be
constructed introducing a set of oscillators for each Heisenberg subalgebra in
the decomposition. These oscillators however could not be independent since the
Heisenberg subalgebras do not commute among themselves. Their commutation
relations, like the ones we calculated in the case of $A_{p-1}$ and $G_2$,
could give hints on how to relate those oscillators. These considerations could
be related to previous attempts to use division algebras, Jordan algebras,
dependent fermions etc, in the construction of representations of Kac-Moody
algebras, see e.g. \cite{jordan}.

We suspect that the orthogonal decomposition presented in section
\ref{sec2} is relevant for studying the completeness problem of the set
of the soliton  solutions to the affine Toda theories obtained in the framework
of the  technique discovered in \cite{OTU93}, especially for their nonabelian
versions discussed in \cite{OSU93}.
Moreover, the orthogonal decomposition being closely related to the
finest gradation of the algebras, could be relevant for a
construction of some multidimensional integrable nonlinear
systems in the framework of the Lie group algebraic approach \cite{ls}.
It would be also interesting to study the analogue of this
decomposition for the algebras of diffeomorphisms groups, as e.g. the
continuous limit of the $A_r$ algebra in the cyclotomic basis to that of the
area preserving diffeomorphisms on a two dimensional manifold. That could be
relevant for some problems related to an affinization of the membrane manifolds
in the context of continuous limits of the gauge theories, 3-d gravity, etc.

\vspace{.5 cm}

{\bf Acknowledgements}

\noindent The authors are grateful to  J.-L. Gervais,
 J. L. Miramontes and J.S\'anchez Guill\'en  for useful discussions;  and to A.
A. Kirillov and A. I. Kostrikin for some important comments.
We also would like to thank the Department of Physics of the
University of Wales, Swansea, where our studies of this
problem  started,  for  kind hospitality, and HEFCW for financial assistance.
M.V.S. was partially supported by the Russian
Fund for Fundamental Research, Grant \# 94-01-01585-A; and ISF, Grant  \# RMO
000; and L.A.F. was partially supported by Ministerio de Educaci\'on y Ciencia
(Spain) and FAPESP (Brazil).


\begin{thebibliography}{**}

\bibitem{KKU81}
A. I. Kostrikin, I. A. Kostrikin and V. A. Ufnarovskii,
{\it Orthogonal Decompositions of Simple Lie Algebras},
Dokl. Akad. Nauk. SSSR {\bf 260}, (1981), 526; English transl. Soviet. Math.
Dokl. {\bf 24}, (1981), 292
\bibitem{KKU811}
A. I. Kostrikin, I. A. Kostrikin and V. A. Ufnarovskii;
{\it Orthogonal Decompositions of Simple Lie Algebras (Type $A_n$)},
Proceedings of the Steklov Institute of Mathematics,  {\bf 4}, (1983),
113-129.
\bibitem{KKU82}
A. I. Kostrikin, I. A. Kostrikin and V. A. Ufnarovskii,
{\it Multiplicative Decompositions of Simple Lie Algebras},
Dokl. Akad. Nauk. SSSR {\bf 262}, (1982); English transl. Soviet. Math.
Dokl. {\bf 25}, (1982), 23-27.
\bibitem{KP85}
V. G. Kac and D. H. Peterson, {\it $112$ constructions of the basic
representation of the loop group of $E_8$},
 In: Symposium on anomalies, geometry and topology. Eds. W.A. Bardeen and A.R.
White, World Scientific (Singapore, 1985).
\bibitem{PZ81}
J. Patera and H. Zassenhaus, {\it The Pauli matrices in $n$ dimensions and the
finest gradings of simple lie algebras of type $A_{n-1}$},
J. Math. Phys. {\bf 29} (3), (1988), 665-673.
\bibitem{kklw} V.G. Kac, D.A. Kazhdan, J. Lepowsky and R.L. Wilson;
{\it Realization of the basic representations of the Euclidean Lie algebras},
Advances in Math. {\bf 42} (1981), 83-112.
\bibitem{kac} V.G. Kac, {\it Infinite Dimensional Lie Algebras}, Third
Edition, Cambridge University Press, 1990.
\bibitem{kostant} B. Kostant, {\it The principal three dimensional subgroup and
the Betti numbers of a complex simple Lie group},
Amer. J. Math. {\bf 81}, (1959), 973-1032.
\bibitem{jordan}
P. Goddard, W. Nahm, D. Olive, H. Ruegg and A. Schwimmer; {\it Fermions and
Octonions}; \CMP{112}{1987}{385-408}.\\
L.A. Ferreira, J.F. Gomes and A.H. Zimerman; {\it Vertex operators and Jordan
Fields}; \PLB{214}{1988}{367-370}; L.A. Ferreira, J.F. Gomes, P. Teot\^onio
Sobrinho and A.H. Zimerman; {\it The Jordan Structure of Lie and Kac-Moody
Algebras}; \JPA{25}{1992}{5071-5088}; {\it Symplectic Bosons, Fermi Fields and
Super Jordan Algebras}; \PLB{234}{1990}{315-320}.\\
E. Corrigan and T. Hollowood; {\it A string construction for a commutative
non-associative algebra related to the exceptional Jordan algebra};
\PLB{203}{1988}{47-51}; {\it The exceptional Jordan algebra and the
superstring}; \CMP{122}{1989}{393-410}.\\
M. Gunaydin and J. Hyun; {\it Affine exceptional Jordan algebra and vertex
operators}; \PLB{209}{1988}{498-502}.
\bibitem{OTU93}
D. I. Olive, N. Turok and J. W. R. Underwood, {\it Solitons and the
energy-momentum tensor for affine Toda theory}, \NPB{401}{1993}{663-697};
{\it Affine Toda Solitons and Vertex Operators}, \NPB{409}{1993}{509-546}.\\
H. Aratyn, C.P. Constantinidis, L.A. Ferreira, J.F. Gomes and A.H. Zimerman;
{\it Hirota's solitons in the affine and the conformal
affine Toda models}, \NPB{406}{1993}{727-770}.
\bibitem{OSU93}
D. I. Olive, M. V. Saveliev and J. W. R. Underwood, {\it On a solitonic
specialisation for the general solutions of some two-dimensional completely
integrable systems}, Phys. Lett. {\bf B311} (1993) 117-122.
\bibitem{ls} A.N. Leznov and M.V. Saveliev, {\it Group Theoretical Methods for
Integration of Nonlinear Dynamical Systems}, Progress in Physics, v. 15,
Birkha\"user-Verlag, Basel, 1992.
\end{thebibliography}
\end{document}